# Intensity mapping without cosmic variance

Trevor M. Oxholm[1,2,*] and Eric R. Switzer[2]

[1]*Department of Physics, University of Wisconsin-Madison, Madison, Wisconsin 53706-1390, USA*
[2]*NASA Goddard Space Flight Center, Greenbelt, Maryland 20771, USA*

Current and future generations of intensity mapping surveys promise dramatic improvements in our understanding of galaxy evolution and large-scale structure. An intensity map provides a census of the cumulative emission from all galaxies in a given region and redshift, including faint objects that are undetectable individually. Furthermore, cross-correlations between line intensity maps and galaxy redshift surveys are sensitive to the line intensity and clustering bias without the limitation of cosmic variance. Using the Fisher information matrix, we derive simple expressions describing sensitivities to the intensity and bias obtainable for cross-correlation surveys, focusing on cosmic variance evasion. Based on these expressions, we conclude that the optimal sensitivity is obtained by matching the survey depth, defined by the ratio of the clustering power spectrum to noise in a given mode, between the two surveys. We find that midinfrared to far-infrared space telescopes could benefit from this technique by cross-correlating with coming galaxy redshift surveys such as those planned for the Nancy Grace Roman Space Telescope, allowing for sensitivities beyond the cosmic variance limit. Our techniques can therefore be applied to survey design and requirements development to maximize the sensitivities of future intensity mapping experiments to tracers of galaxy evolution and large-scale structure cosmology.

## I. INTRODUCTION

Intensity mapping is an emerging approach to studying galaxy evolution and large-scale structure cosmology. Intensity maps trace the cumulative emission of a given line from all galaxies within a region, making them sensitive to the line luminosity function without the selection effects present in direct-detection surveys. Because of modest aperture requirements and advancements in detector and spectrometer technologies, intensity mapping surveys can be used to economically map extremely large volumes of the Universe.

While the present generation of intensity mapping surveys are limited by instrument noise, coming generations will be limited by cosmic variance [1–3]. Cosmic variance is intrinsic to measurements of the power spectrum of a fluctuating field and can only be improved by measuring a larger number of Fourier modes. Furthermore, variance in the matter density field leads to cosmic variance in the inference of the line intensity and clustering bias, owing to a degeneracy between the three parameters.

Cosmic variance can be avoided by cross-correlating the intensity map with another large-scale structure survey tracing the same matter density field [4–7]. Information from the cross-correlating survey can be used to break the degeneracy between the intensity and the matter overdensity. Furthermore, cross-correlation with a galaxy redshift survey is established as an approach to reject contamination in intensity maps from foregrounds and interloping lines, which do not correlate with emission from galaxy redshift survey population [8–10]. These sources of uncorrelated variance contribute to error bars in the cross-correlation but do not additively bias the result, as they would for intensity autocorrelations.

The formalism for cosmic variance evasion through cross-correlations has been well-developed in the context of weak lensing surveys [11]. Previous studies describe cosmic variance-free measurements of structure growth through redshift-space distortions [12] and primordial non-Gaussianity through scale-dependent biases [13]. While these studies provide a useful guide for intensity mapping cross-correlations, intensity maps are unique in their dependence on the line mass-luminosity function, as well as correlated shot noise between the two surveys. These unique features can be exploited to extend the cosmic variance evasion technique to measurements of galaxy evolution. Furthermore, the biased intensity can be extended to include scale-dependent bias and redshift-space distortions, thereby providing cosmic variance-free measurements of $f_{\rm NL}$ and the growth factor, respectively, as studied in Bernstein and Cai [12], Seljak [13].

Analysis of cross-correlation sensitivities may also be used to devise optimal survey strategies, in terms of survey depth or integration time per pixel. Cosmic microwave background (CMB) maps are analogous to intensity maps at discrete redshifts without the inclusion of a cross-correlating survey. Here, the error per mode is proportional to a cosmic variance term plus an instrument noise term, which decreases with increased survey depth. It is therefore optimal to increase the survey depth until cosmic variance provides the dominant source of variance, before focusing on an increasing survey area to detect more modes; this conclusion matches that of Knox [14] regarding CMB sensitivities. For intensity mapping cross-correlations, on the other hand, variance per mode is optimized when the depths are matched between the intensity map and the cross-correlating map. Further details on the formalism are described in Sec. III.

[*]oxholm@wisc.edu

We focus on intensity maps of singly ionized carbon ([CII]). [CII] is of particular interest to studies of galaxy evolution, owing to its reliable tracing of the star formation rate density and high brightness among cooling lines [15]. Furthermore, [CII] intensity maps may also provide powerful measurements of $f_{\rm NL}$ [16]. Pullen *et al.* [17] and Yang *et al.* [10] obtain preliminary detections of the [CII] mean intensity at $z = 2.6$. Dedicated experiments aim to follow up these measurements over a broad redshift range, including CCAT-prime [18], Concerto [19], EXCLAIM [20], TIM [21], and TIME [22]. Proposed next-generation space telescopes such as PIXIE [1,23], GEP [24], and the Origins Space Telescope [25] may dramatically improve intensity mapping through far-infrared to midinfrared lines, owing to the background-limited sensitivity of space platforms. We will focus on a realistic but generic space background-limited telescope for our forecasts.

While we study cross-correlations between [CII] intensity maps and galaxy redshift surveys, our techniques can be applied to any other intensity mapping cross-correlations, including those between two intensity maps. Particularly, this formalism can be applied to survey design and requirements development for future surveys limited by cosmic variance, analogous to the Knox formula [14] describing CMB fluctuations. Furthermore, we show that these calculations are robust to additive biases in the intensity mapping power spectrum, which may be attributed to foregrounds or interloping lines.

We build on existing formalism from Switzer *et al.* [6] and Bernstein and Cai [12] for intensity mapping cross-correlations featuring correlated shot noise. We derive a simple expression for the cross-correlation sensitivity and generalize the results to a variety of scenarios. We focus on cross-correlation between [CII] and a galaxy redshift survey. Finally, we describe optimal survey strategies based on our results.

## II. INTENSITY MAPPING CROSS-CORRELATIONS

### A. Tracing the matter density

The primary observable in an intensity map is the intensity field $\delta_I$, which traces large-scale matter fluctuations $\delta_m$ as

$$\delta_I(\mathbf{r}, z) = I(z) b(z) \delta_m(\mathbf{r}, z), \tag{1}$$

where $\mathbf{r}$ represents spatial location and $b$ is the linear clustering bias of the observed galaxies relative to the underlying matter overdensity $\delta_m$. $I(z)$ is the intensity at mean density, representing the cumulative emission from all galaxies in the survey region.

The intensity $I$ is related to the first moment of the galaxy luminosity function, and its dependence on $z$ probes galaxy evolution. Using the halo model approach [26], we calculate

$$I(z) = \alpha(z) \int_{M_{\rm min}}^{\infty} L(M, z) n(M, z) dM, \tag{2}$$

where $M$ is the halo mass and $n(M, z)$ is the halo mass function. $L(M, z)$ is the mass-luminosity relation for the given line and is the primary observable that constrains galaxy evolution physics. $M_{\rm min}$ the minimum mass of a halo hosting a galaxy with the target line. Lower-mass halos tend to feature suppressed star formation activity, resulting in a negligible contribution to the mean line intensity [27]. We impose a strict halo mass cutoff of $M_{\rm min} = 10^{10}\, M_\odot$ matching cosmic infrared background models [28], though Eq. (2) is not sensitive to order-of-magnitude changes in the halo mass cutoff, consistent with conclusions within Shang *et al.* [29]. Because we focus on large-scale clustering, our conclusions are not sensitive to changes in the halo mass profile that may be present on smaller scales, which may be sensitive to $M_{\rm min}$ through the halo occupation distribution.

The prefactor $\alpha(z)$ is given by

$$\alpha(z) = \frac{1}{4\pi \nu_{\rm rest} H(z)}, \tag{3}$$

accounting for the conversion from integrated luminosity to intensity (i.e., units of $L_\odot$ to $\mathrm{Jy\,sr^{-1}\,h^{-3}\,Mpc^3}$), as well as luminosity distance and angular diameter distance effects. Here, $\nu_{\rm rest}$ is the rest frame frequency of the target emission, and $H(z)$ is the Hubble parameter at redshift $z$.

The bias $b$ between the intensity field and the underlying matter density field provides another observable that may be used to infer the mass-luminosity function. We calculate $b$ as

$$b(z) = \frac{\int_{M_{\rm min}} dM b_h(M, z) L(M, z) n(M, z)}{\int_{M_{\rm min}} dM L(M, z) n(M, z)}, \tag{4}$$

where $b_h(M, z)$ is the bias of a given halo against the underlying matter power spectrum [30,31].

The intensity and bias are completely degenerate, so we group the terms together as the *biased intensity* $[Ib]$. In the calculations that follow, we define the biased intensity simply as the multiplication between the intensity $I$ and a constant bias $b$, specified by Eqs. (2) and (4), without the inclusion of redshift-space distortions (RSD) or scale-dependent bias. $[Ib]$ therefore contains information about the integral of the line mass-luminosity function $L(M,z)$ and the halo mass cutoff $M_{\rm min}$. However, our formalism can be generalized to include RSD and scale-dependent biases in $[Ib]$ without significantly changing our conclusions. This generalization would provide information on structure growth, primordial non-Gaussianity, and nonlinear effects leading to scale-dependent biases on small scales. The mass of halos hosting a star formation could also be measured through $M_{\rm min}$, benefiting from the use of RSD to separate $I$ from $b$.

Using only information from the intensity map, $[Ib]$ is also degenerate with $\delta_m$. The latter is subject to cosmic variance because it depends on the measurement of waves in the matter density field, with a variance limited by the number of observable modes. Because $[Ib]$ and $\delta_m$ are degenerate, cosmic variance in $\delta_m$ is coupled to variance in $[Ib]$.

The degeneracy may be broken through the inclusion of a second map tracing the matter overdensity, e.g., a galaxy redshift survey. In this case, the galaxy field $\delta_g$ is given by

$$\delta_g(\mathbf{r},z) = b_g(z)\delta_m(\mathbf{r},z), \tag{5}$$

where the bias $b_g$ pertains to galaxies observed in the galaxy redshift survey. Generally, $b \neq b_g$ because intensity mapping and galaxy redshift surveys do not observe the same galaxy populations. This can be attributed to different halo mass cutoffs for the two tracers (set by luminosity cutoffs in galaxy redshift surveys) and scatter in the probability a given halo will be populated by line-emitting gas or a directly observable galaxy.

Figure 1 shows a comparison between the three realistic simulated fields $\delta_m$, $\delta_I$, and $\delta_g$. Because $\delta_g$ traces $\delta_m$, we can use it to measure the fluctuations in the density field, allowing us to remove the contributions from $\delta_m$ to cosmic variance in $\delta_I$.

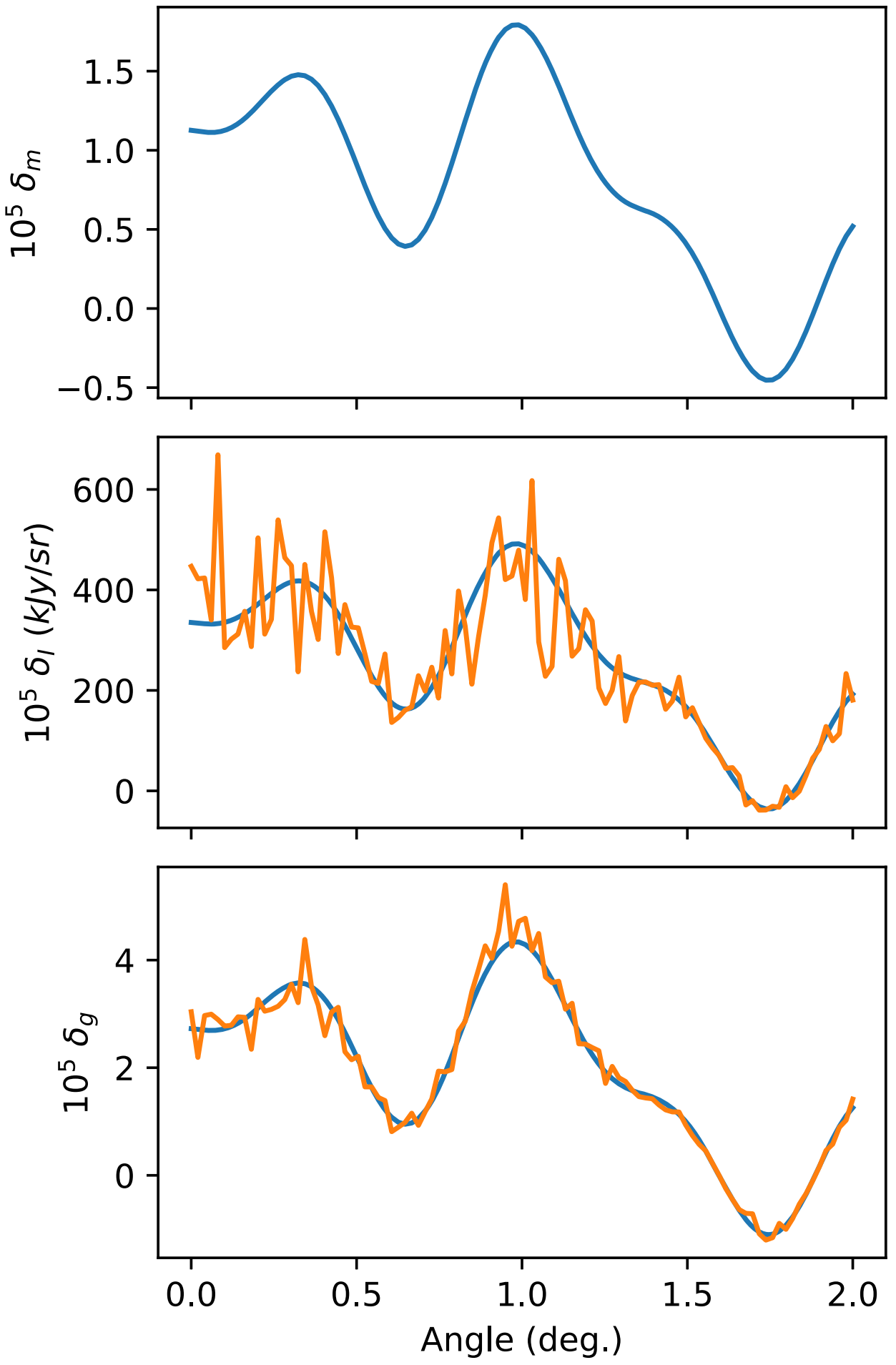

FIG. 1. Fluctuations in the overdensity of the matter ($\delta_m$), intensity ($\delta_I$), and galaxy ($\delta_g$) fields. $\delta_I$ and $\delta_g$ trace the same matter overdensity $\delta_m$ according to Eqs. (1) and (5). The fields in the lower two panels are shown without noise (blue curves), and with fractional noise of 30% and 10% in $\delta_I$ and $\delta_g$, respectively. Note that $\delta_m$ was constructed by superimposing 10 randomly selected $k$ modes of random phase, with amplitudes corresponding to amplitudes in the matter power spectrum.

### B. Cross-correlation statistics

The power spectrum of the intensity map is given by

$$P_I(\mathbf{k},z) = T_I^2(\mathbf{k},z)([Ib]^2(z)P_m(k,z) + I_2(z)) + P_N(z), \tag{6}$$

where $P_m(z)$ is the matter power spectrum describing fluctuations in $\delta_m$ [32], $\mathbf{k}$ is the wave vector, and the wave number $k = |\mathbf{k}|$. The first (clustering) term is proportional to the square of the first moment of the luminosity function, along with a bias $b$. $I_2(z)$ is the shot power of the intensity map and is proportional to the second moment of the galaxy luminosity function [33]:

$$I_2(z) = \alpha^2(z)\int_{M_{\rm min}}^{\infty} L^2(M,z)n(M,z)dM. \tag{7}$$

$P_N$ represents the instrument noise, given by $P_N = \sigma_N^2 V_{\rm vox}$ with $\sigma_N^2 = \mathrm{NEI}^2/t_{\rm int}$.

NEI is the noise-equivalent intensity given in units of $\mathrm{Jy\,sr^{-1}\,s^{1/2}}$, $V_{\rm vox}$ is the voxel volume, and $t_{\rm int}$ is the integration time per pixel. The resolution transfer function, $T_I$, accounts for the loss of information on small scales due to finite spatial and frequency resolution in the intensity mapping instrument. We model $T_I$ using a Gaussian profile, specified identically by, e.g., Eq. (13) in Bernal *et al.* [34].

For later convenience, we reparameterize Eq. (6) in terms of a dimensionless noise term $W_I$ given by

$$W_I(\mathbf{k},z) = \frac{I_2(z) + P_N(z)/T_I^2(\mathbf{k},z)}{[Ib]^2(z)P_m(k,z)}, \tag{8}$$

representing fractional noise in the intensity mapping measurement of the clustering power spectrum due to shot noise and instrument noise. $W_I$ can be generalized to also include residual foregrounds and interloping lines. With this substitution, Equation (6) evaluates to

$$P_I(\mathbf{k},z) = [Ib]^2(z)T_I^2(\mathbf{k},z)P_m(k,z)[1 + W_I(\mathbf{k},z)]. \tag{9}$$

We can analogously define the power spectra associated with the galaxy redshift survey, $P_g$, and the cross-correlation between the two surveys $P_\times$ as

$$P_g(k,z) = b_g^2 P_m[1 + W_g(k,z)], \tag{10}$$

$$P_\times(\mathbf{k},z) = T_I(\mathbf{k},z)[Ib]b_g P_m[1 + W_\times(k,z)], \tag{11}$$

with associated dimensionless noise terms given by

$$W_g(k,z) = \frac{1}{\bar{n}(z)b_g^2(z)P_m(k,z)}, \tag{12}$$

$$W_\times(k,z) = \frac{P_{\mathrm{shot}}^\times(z)}{[Ib](z)b_g(z)P_m(k,z)}. \tag{13}$$

Here, $\bar{n}$ represents the number density of galaxies observed in the galaxy redshift survey and $P_{\mathrm{shot}}^\times$ is the cross-shot power, representing correlated shot noise from the number of galaxies observed by both surveys. We estimate the cross-shot power as [33]

$$P_{\mathrm{shot}}^\times(z) = \alpha(z)\int_{M_{\mathrm{min}}^\times}^\infty L(M,z)n_g(M,z)n(M,z)dM, \tag{14}$$

where $n_g(M)$ is the halo occupation distribution of the galaxy redshift survey, which contributes to the number density of detected galaxies as $\bar{n} = \int_{M_{\mathrm{min}}^g}^\infty n_g(M,z)n(M,z)dM$, with $M_{\mathrm{min}}^g$ the halo mass corresponding to the lower-luminosity threshold. $M_{\mathrm{min}}^\times$ is the minimum halo mass for galaxies observable in both surveys; we take $M_{\mathrm{min}}^\times = M_{\mathrm{min}}^g$ assuming the luminosity threshold in the galaxy redshift survey does not reach the smallest halos contributing to the [CII] intensity. Note that Eq. (14) is approximate, as we have left out potentially important physics, such as scatter between the two galaxy populations, a soft cutoff for $M_{\mathrm{min}}^\times$, and contributions from the one-halo power spectrum. Later in the paper, we will explore marginalizing over $P_{\mathrm{shot}}^\times$, leading to increased variance on small scales where the shot power is non-negligible compared to the clustering power spectrum.

We stress that the dimensionless noise parameters $W_I$, $W_\times$, and $W_g$ are scale dependent; on large scales where the clustering power spectra tend to dominate, the parameters tend to be much smaller than one. On small scales where instrument noise, shot power, and/or the one-halo power spectrum dominate, the dimensionless noise parameters are larger than one, signifying increased variance in the maps.

### C. Autocorrelation and cross-correlation sensitivities

We derive the sensitivity of the cross-correlation using the Fisher matrix formalism. We begin with the assumption that we can utilize information from all three maps: the intensity map, the galaxy map, and the cross-correlation map. In this case, the covariance matrix of the intensity map and galaxy redshift survey is given by

$$\Sigma = \begin{pmatrix} P_I & P_\times \\ P_\times & P_g \end{pmatrix}, \tag{15}$$

and the Fisher matrix is given by Tegmark *et al.* [35],

$$F_{ij} = \frac{1}{2}\mathrm{Tr}\left[\Sigma^{-1}\frac{\partial\Sigma}{\partial p_i}\Sigma^{-1}\frac{\partial\Sigma}{\partial p_j}\right], \tag{16}$$

where $p_i$ corresponds to the varied parameters, e.g., $[Ib]$ and $P_m$.

As a first step, we calculate errors assuming information only in the line intensity map, i.e., the line intensity autocorrelation. Setting $\Sigma = P_I$ and inserting into Eq. (16) with $p_i = [Ib]$, we arrive at the autopower errors,

$$\left.\frac{\sigma_{Ib}^2(\mathbf{k})}{(Ib)^2}\right|_{\mathrm{auto}} = \frac{(1 + W_I(\mathbf{k}))^2}{2N_{\mathrm{modes}}(\mathbf{k})}, \tag{17}$$

with $N_{\mathrm{modes}}$ the number of modes at a given wave number $k$. Note that $z$ dependence is implied. The first term in the parentheses describes the cosmic variance limit while the second describes variance in the map not originating from clustering of large-scale structure. Equation (17) is analogous to the Knox formula describing errors in the cosmic microwave background [14], and the Feldman-Kaiser-Peacock formula for errors in galaxy redshift surveys [36].

We can estimate the number of modes for a given wave number $k$ and line of sight angle $\mu$ by using the isotropic mode-counting formula,

$$N_{\mathrm{modes}}(\mathbf{k}) \approx \frac{k^2 V_{\mathrm{surv}}}{8\pi^2}\delta k\delta\mu, \tag{18}$$

where $\mu$ is the angle of $\mathbf{k}$ with respect to the line of sight, $\delta k$ and $\delta\mu$ are the widths of the $k$ and $\mu$ bins, and $V_{\mathrm{surv}}$ is the survey volume.

As a next step, we calculate errors in the inference of $[Ib]$ using information between both $P_I$ and $P_g$, and the cross-correlation $P_\times$. Inserting the full covariance matrix [Eq. (15)] into the Fisher matrix, marginalizing over $P_m$, and varying over $[Ib]$, we arrive at the exact solution

$$\left.\frac{\sigma^2_{Ib}(\mathbf{k})}{[Ib]^2}\right|_{\times} = \frac{W_I - 2W_\times + W_g + W_I W_g - W^2_\times}{N_{\rm modes}(\mathbf{k})}. \qquad (19)$$

The cosmic variance term is no longer present, while there are additional contributions from $W_\times$ and $W_g$. Note that $W_\times$ contributes negatively to the fractional errors; this feature was pointed out by Liu and Breysse [7] studying constraints on $f_{\rm NL}$ using intensity mapping cross-power spectra. This effect can be contextualized through the map cross-correlation coefficient $\mathcal{R}_\times = (1 + W_\times)/\sqrt{(1 + W_I)(1 + W_g)}$. Cross-correlation errors are generally minimized when $\mathcal{R}_\times \to 1$, so increased correlated shot noise leads to decreased variance.

Finally, we stress that the variances listed in this section correspond to specific $\mathbf{k}$ vectors, specified by a magnitude k and the cosine of the angle to the line-of-sight $\mu$. Accurate variance per k must be performed by integrating Fisher matrices over $\mu$, i.e., $F_{ij}(k) = \int d\mu F_{ij}(k, \mu)$. This is particularly important when factoring in RSD, and in modes where the resolution transfer function approaches zero. The $\mu$ dependence enters the resolution transfer function because generally the beam and frequency resolutions do not yield identical limits on perpendicular and parallel modes, respectively.

Note that while we focus on cross-correlation with galaxy redshift surveys, our formalism can be extended to other cross-correlations, including those between two intensity maps. However, this requires an understanding of correlated noise between the two maps and requires further study.

### D. Robustness of the cross-correlation sensitivity

The qualitative features of the errors described by Eq. (19) are robust to a variety of assumptions on the model and availability of data, namely, marginalizing over the cross-shot power, the noninclusion of intensity mapping autocorrelation data, and additional sources of stochasticity in the cross-power spectrum.

#### *1. Marginalizing over shot power and/or one-halo terms in the cross-power spectrum*

Our formalism for the cross-shot power given in Eq. (14) is a rough estimate and leaves out many important details related to astrophysics and the one-halo power spectrum. We therefore examine the conservative approach of marginalizing over $P^\times_{\rm shot}$. Marginalizing over $W_\times$ and $P_m$, Eq. (19) instead evaluates exactly to

$$\left.\frac{\sigma^2_{Ib}(\mathbf{k})}{[Ib]^2}\right|^{\rm marg.}_{P^\times_{\rm shot}} = \frac{W_I - 2W_\times + W_g + \frac{1}{2}(W_I^2 + W_g^2) - W^2_\times}{N_{\rm modes}(\mathbf{k})}, \qquad (20)$$

$$\left.\frac{\sigma^2_{P^\times_{\rm shot}}(\mathbf{k})}{P^2_{\times\rm shot}}\right|^{\rm marg.}_{Ib} = \frac{2 - 2\tilde{W} + \tilde{W}^2/2}{N_{\rm modes}(\mathbf{k})}, \qquad (21)$$

where we have also solved for errors in the cross-shot power for completeness, and we have defined $\tilde{W} = (W_I + W_g)/W_\times$ for brevity. Note that here cosmic variance contributes to the cross-power spectrum errors as a factor of $2/N_{\rm modes}$.

Comparing Eq. (20) to Eq. (19), we find that terms linear in the $W$ parameters are identical. We refer to these terms linear in $W$ as the first-order errors. Higher-order terms, however, are different between the two surveys; terms quadratic in $W_I$ and $W_g$ appear. On small scales where the clustering power spectrum is subdominant in either the intensity or galaxy redshift survey maps ($W_I$ or $W_g > 1$), marginalizing over $P^\times_{\rm shot}$ results in higher errors. On large scales where $W_I$ and $W_g > 1$, marginalizing over $P^\times_{\rm shot}$ has little effect on the variance.

#### *2. Exclusion of information from the autopower of the intensity map*

The covariance matrix in Eq. (15) assumes that information is usable in both the intensity map and the galaxy redshift survey. In many cases, the intensity mapping autocorrelation signal can be contaminated by additive biases due to foregrounds, interloping lines, and other systematics. It is therefore useful to calculate errors assuming no information from the intensity map autocorrelation.

In this case, we define the data vector $\Theta$ by the two-point correlation functions defined by $\Theta = (P_\times, P_g)$. The covariance matrix is given by

$$\Sigma = \begin{pmatrix} C^{Ig}_{Ig} & C^{gg}_{Ig} \\ C^{gg}_{Ig} & C^{gg}_{gg} \end{pmatrix}, \qquad (22)$$

where $C^{Ig}_{Ig} = P^2_\times + P_I P_g$, $C^{gg}_{Ig} = 2P_\times P_g$, and $C^{gg}_{gg} = 2P^2_g$ are the covariances per mode, assuming Gaussian fields in $\delta_I$ and $\delta_g$.

Using the Fisher matrix for information in the mean (rather than the covariance),

$$F_{ij} = \frac{1}{2}\mathrm{Tr}\left[\Sigma^{-1}\left(\frac{\partial\Theta^T}{\partial p_i}\frac{\partial\Theta}{\partial p_j} + \frac{\partial\Theta^T}{\partial p_j}\frac{\partial\Theta}{\partial p_i}\right)\right], \qquad (23)$$

and plugging Eq. (21) and $\Theta$ into Eq. (23), we again find that the first-order sensitivity in $[Ib]$ is identical to that shown in Eq. (19), with differences appearing at second order:

$$\left.\frac{\sigma^2_{Ib}(\mathbf{k})}{[Ib]^2}\right|_{\times} = \frac{W_I - 2W_\times + W_g}{N_{\rm modes}} + \frac{W_I W_g - 4W_\times W_g + W^2_\times + 2W^2_g}{N_{\rm modes}}. \qquad (24)$$

Cosmic variance is still suppressed because $P_m$ can be marginalized over through $P_\times$ and $P_g$.

For a final case, we calculate errors including only information from the cross-correlation $P_\times$. Using a covariance $\Sigma = C_{Ig}^{Ig}$ and data $\Theta = P_\times$ and plugging into Eq. (23), we arrive at

$$\left.\frac{\sigma_{Ib}^2(\mathbf{k})}{[Ib]^2}\right|_{\times\text{only}} = \frac{1}{N_\text{modes}} + \frac{(1+W_I)(1+W_g)}{N_\text{modes}(1+W_\times)^2}. \quad (25)$$

Here, cosmic variance has reappeared because it is not possible to marginalize over $P_m$.

#### 3. Additional sources of stochasticity

The galaxy redshift survey and intensity mapping survey may not identically trace the underlying matter field. We benchmark this effect through a cross-correlation (stochasticity) coefficient $r_\times$, contributing to the cross-power spectrum as

$$P_\times^r = [Ib]b_g P_m[r_\times + W_\times]. \quad (26)$$

This is purely a choice of convention, as all sources of decorrelation can also be absorbed into $W_\times$ and marginalized or measured. Marginalizing over $W_\times$ and $P_m$, residual stochasticity contributes to the fractional variance in the cross power [analogous to Eq. (20)] as

$$\left.\frac{\sigma_{Ib}^2(\mathbf{k})}{[Ib]^2}\right|_{r_\times} = \frac{1 - r_\times^2 - 2r_\times W_\times + W_I + W_g}{N_\text{modes}(\mathbf{k})} + \frac{-W_\times^2 + (W_I^2 + W_g^2)/2}{N_\text{modes}(\mathbf{k})}. \quad (27)$$

Note that cosmic variance appears when the cross-correlation coefficient $r_\times \neq 1$, and Eq. (20) is recovered in the limit that $r_\times \to 1$. In the limit that $r_\times \to 0$, we recover a similar cosmic variance term to Eq. (17). Note that nonlinear processes that cause $r_\times < 1$ tend to appear on small scales, similar to those where the shot power becomes dominant; effects include scale-dependent biases and contributions to the one-halo power spectrum that may be incorrectly modeled.

Equation (27) is useful to illustrate the importance of highly correlated maps in canceling cosmic variance. Figure 2 shows error contours for a single mode in the $[Ib] - P_m$ plane for two cases: perfectly uncorrelated maps ($r_\times = 0$), and perfectly correlated maps ($r_\times = 1$). In the uncorrelated case, the degeneracy between $P_m$ and $[Ib]$ is shown clearly in the error ellipse; cosmic variance errors in $P_m$ are coupled to errors in $[Ib]$. Note, however, that the degeneracy is mildly broken by the appearance of $P_m$ in the clustering power of the galaxy redshift survey. In the perfectly correlated case, the degeneracy is broken and errors in $[Ib]$ are no longer coupled to those in $P_m$, so errors in the former are limited by noise in the maps, rather than cosmic variance.

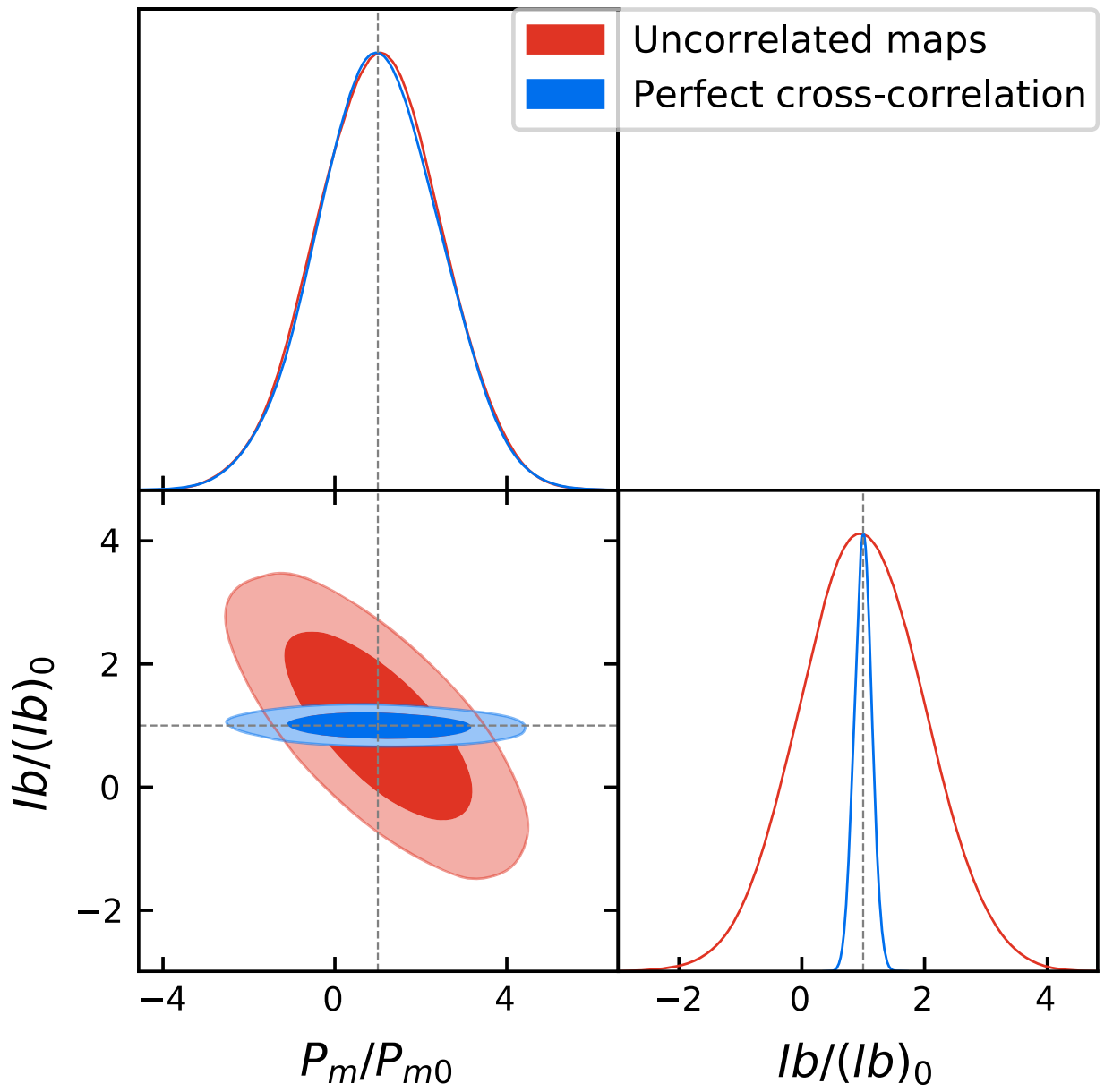

FIG. 2. The cosmic variance evasion technique for a single mode in the cross-power spectra given by Eq. (27) with noise parameters arbitrarily chosen as $(W_I, W_\times, W_g) = (0.01, 0, 0.01)$. The red ellipse shows the case where $r_\times = 0$, i.e., the maps are uncorrelated, resulting in a variance-limited measurement in $[Ib]$. Here, measurements of $Ib/(Ib)_0$ and $P_m/P_{m0}$ are degenerate in $P_I$, resulting in coupling between errors in the $P_m$ and $[Ib]$. The blue ellipse shows the $r_\times = 1$ case, where we are able to flatten the covariance ellipse in the $Ib/(Ib)_0 - P_m/P_{m0}$ plane, allowing us to marginalize the $P_m$ errors with minimal residual errors.

## III. DESIGNING AN INTENSITY MAPPING SURVEY WITHOUT COSMIC VARIANCE

### A. Survey optimization

Given a fixed total time for a survey, Eqs. (17) and (19) can be used to optimize the total survey area $A_\text{surv}$, in order to reduce errors in a given $k$. Throughout this section we assume $W_I$ is dominated by instrument noise, rather than shot noise or residual foregrounds. Increased integration time per pixel linearly decreases instrument noise and therefore $W_I$. However, the increased integration time per pixel results in a smaller number of observed pixels, linearly decreasing the scan area $A_\text{surv}$ and therefore $N_\text{modes}$. Different terms present in Eqs. (17) and (19) scale with survey area (with fixed total survey time) as

$$\frac{1}{N_\text{modes}} \propto A_\text{surv}^{-1}, \quad \frac{W_I}{N_\text{modes}} \propto A_\text{surv}^{0}, \quad \frac{W_I^2}{N_\text{modes}} \propto A_\text{surv}^{1}. \quad (28)$$

Errors in the autocorrelation scale analogously to conclusions made in Knox [14]. When instrument noise dominates, there is a clear preference for small survey coverage with increased integration time per pixel. This is because, for an instrument noise-dominated survey, autocorrelation errors are proportional to $W_I^2/N_{\rm modes}$, which increase linearly with $A_{\rm surv}$. When $W_I < 1$, i.e., when cosmic variance dominates the errors, it is best to widen the survey area. We therefore suggest optimizing the integration time per pixel for autocorrelation so that $W_I$ is just below 1, over the widest area possible.

Cross-correlation errors scale slightly differently from autocorrelation errors. When we do not marginalize over $W_\times$, Eq. (19) shows that $W_I$ does not contribute beyond linear order. Based on the scaling of $W_I$ with $N_{\rm modes}$, this would suggest no preference for wide or deep surveys when instrument noise dominates, but again, a clear preference for wide surveys when $W_I < 1$. On the other hand, if we marginalize over the cross-shot power, Eq. (20) shows that the variance is quadratic in $W_I$, increasing proportionally to $A_{\rm surv}$; thus, a deep survey is preferred. Generally, however, deep surveys emphasize high-$k$ modes, putting more weight on the modes where correlated shot noise is marginalized.

In the case where instrument noise is low, errors in the cross power are not cosmic variance limited. Instead, errors are limited by $W_g$, as described in Eqs. (19) and (20). Therefore, it is ideal to set the integration time such that the depths of the two surveys match, i.e., $W_I \approx W_g$. Increasing the integration time beyond this limit would only decrease $W_I$ while $W_g$ remains constant; increasing $A_{\rm surv}$, on the other hand, would increase $N_{\rm modes}$, thereby decreasing errors attributed to both $W_I$ and $W_g$. This survey strategy extends the arguments of Knox [14] to the cross power; errors in the cross-correlation are limited by noise in the galaxy redshift survey, rather than cosmic variance.

### B. Realistic survey model

We benchmark our model by producing a realistic intensity mapping cross-correlation strategy for a realistic cross-correlation between a space-based [CII] line-intensity mapping survey and the nominal 2000 deg$^2$ High-Latitude Survey for the Nancy Grace Roman Space Telescope [37] at $z = 1.5$.

The rest frequency of [CII] is $\nu_{\rm CII}^{\rm rest} = 1900$ GHz. We calculate the [CII] intensity by inserting the mass-luminosity function of Padmanabhan [38] into Eqs. (2) and (4), resulting in a biased intensity of $[Ib](z = 1.5) = 144\ {\rm Jy\,sr^{-1}}$. We model the Roman HOD through HOD0 in [39], with a multiplicative constant scaled to obtain an average galaxy number density of $\bar{n}(z = 1.5) = 0.0016\ {\rm h^3\,Mpc^{-3}}$, and we calculate a bias $b_g(z = 1.5) = 2.3$.

We assume a $1/e$ beamwidth of 1 arcmin corresponding to a 70 cm primary mirror, and a spectral resolving power $R = 500$, resulting in a frequency resolution $\delta\nu = 1.27$ GHz, which we take to be equal to the single-channel bandwidth.

We assume the noise in the intensity map is background limited with a background intensity of $I_{\rm bknd} = 9\times 10^3\ {\rm kJy\,sr^{-1}}$, an accurate figure for the combined CMB, galactic Cirrus in typical clean regions out of the galactic plane, and CIB radiation [40–42]. Our strategy to calculate the NEI is to first relate $I_{\rm bknd}$ to the power absorbed by the detectors $P_{\rm det}$, through the relation $P_{\rm det} = \eta_{\rm det} I_{\rm bknd} \partial P/\partial I$, with $\eta_{\rm det}$ the detector efficiency and $\partial P/\partial I = A_d \Omega_{\rm inst} \delta\nu$, assuming an unpolarized source.

Here, the detector area $A_d$ and the beam area $\Omega_{\rm inst}$ are related to the frequency $\nu = 760$ GHz and number of optical modes $N_{\rm opt}$ as $A_d \Omega_{\rm inst} = N_{\rm opt} \eta_{\rm opt} (c/\nu)^2$, with $\eta_{\rm opt}$ the optical efficiency. We assume $N_{\rm opt} = 1$ and $\eta_{\rm opt}\eta_{\rm det} = 0.20$. The resulting detector power is translated into a photon background-limited noise-equivalent power ${\rm NEP} = [h\nu P_{\rm det} + P_{\rm det}^2/\delta\nu]^{1/2}$ (e.g., [43]), where $h$ is Planck's constant. This results in ${\rm NEP} = 4.6 \times 10^{-20}\ {\rm W\,Hz^{-1/2}}$, a realistic goal for, e.g., far-infrared kinetic inductance detectors [44,45]. Finally, we calculate ${\rm NEI} = {\rm NEP}/\eta_{\rm det}/(\sqrt{2}\partial P/\partial I) = 69.2\ {\rm kJy\,sr^{-1}}$. The factor of $\sqrt{2}$ comes from the conversion from power spectral conventions of ${\rm Hz}^{1/2}$ to ${\rm s}^{-1/2}$. The model survey occupies 2000 deg$^2$ over a range of redshift $\Delta z = 0.2$ over 417 days (5 s per $1/e$ beamwidth), resulting in $V_{\rm surv} = 1.4\ {\rm h^{-3}\,Gpc^3}$.

The resulting power spectra are shown in Fig. 3, and sensitivities are shown in Fig. 4. The latter shows the fractional variance per $k$ given three cases: autopower, cross-power errors using information from the cross power only (labeled CV), and cross-power errors using information from all available maps, marginalized over $P_m$ and $P_\times^{\rm shot}$; these errors are calculated by Eqs. (17), (25), and (20), respectively. We marginalize over $P_\times^{\rm shot}$ in the latter case to remain conservative about our simplified model describing Eq. (14). We assume 32 band powers in $k$ spanning $0.01 < k < 2.7\ {\rm h\,Mpc^{-1}}$ equally spaced logarithmically, with Fisher matrices integrated over $\mu$. We note that instrument noise dominates shot power, and we assume sufficient bandpass stability that noise due to residual foregrounds is also subdominant.

The cosmic variance-evading cross power significantly outperforms the autopower and cross (CV) errors on large scales where the intensity and galaxy redshift survey maps are highly correlated, i.e., $\mathcal{R}_\times \approx 1$. On smaller scales ($k \gtrsim 0.3\ {\rm h\,Mpc^{-1}}$), autopower errors outperform cross-power errors because dimensionless noise in the galaxy redshift survey exceeds cosmic variance, i.e., $W_g > 1$. The CV case also outperforms the marginalized cross power on smaller scales ($k \gtrsim 1.1\ {\rm h\,Mpc^{-1}}$) because the latter is marginalized over the $P_{\rm shot}^\times$, thereby increasing errors.

Figure 5 shows variance per $k$ as a function of the survey area, assuming the same total survey time. We calculate the variance for two modes: $k = 0.02$ and $0.80\ {\rm h\,Mpc^{-1}}$. Variance in the $0.02\ {\rm h\,Mpc^{-1}}$ mode is minimized by increasing $A_{\rm surv}$ as much as possible. This is because $W_I < W_g$ even for the largest survey area, so increased survey depth does little to decrease variance in $[Ib]$. Furthermore, on these large scales the cross power provides a drastic increase in sensitivity compared to the cosmic variance-limited autopower.

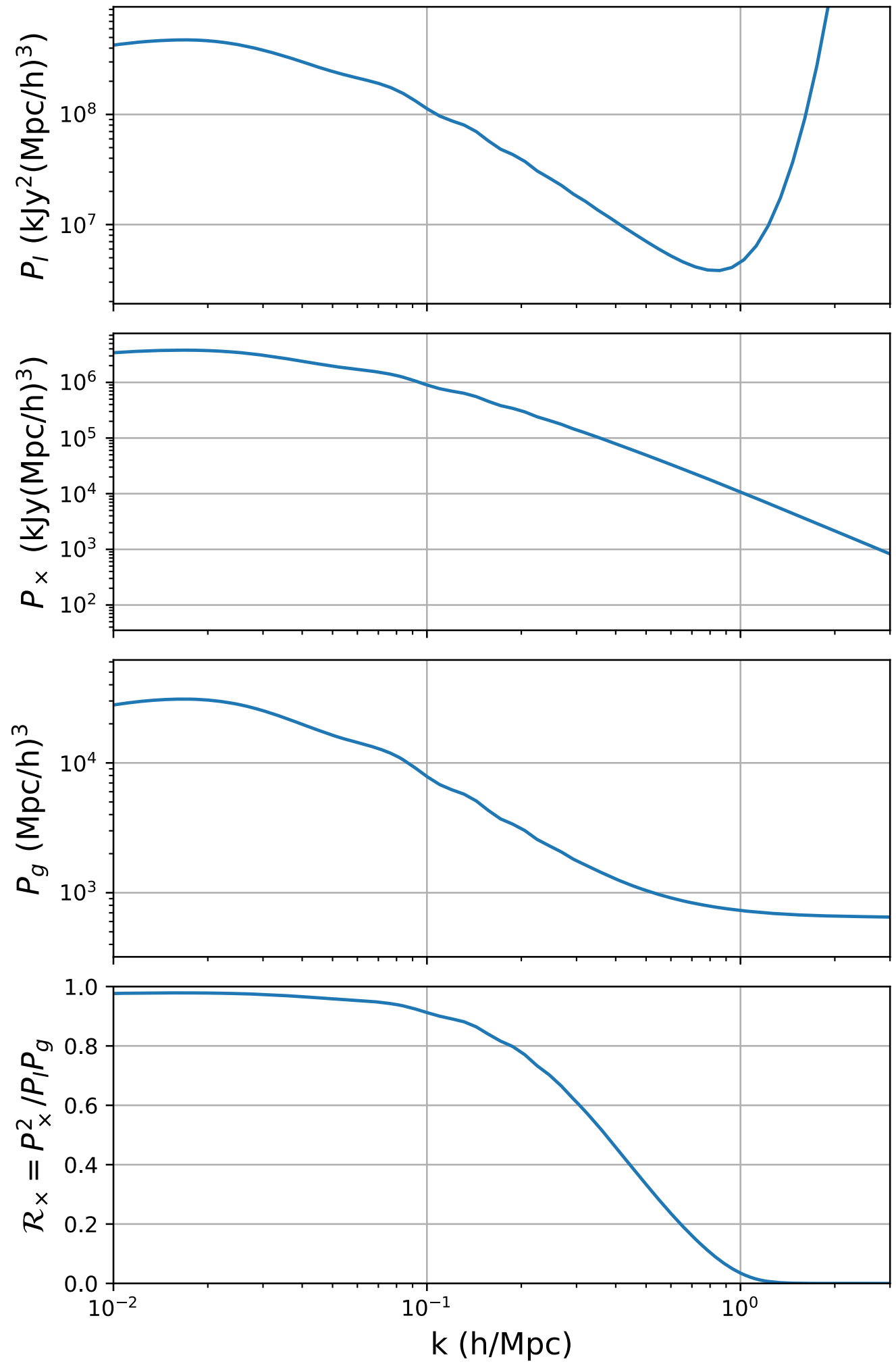

FIG. 3. Power spectra corresponding to the wide survey model in III B: intensity mapping auto-power spectrum (top), cross-power spectrum (upper middle), galaxy auto-power spectrum (lower middle), and map-space cross-correlation coefficient (bottom).

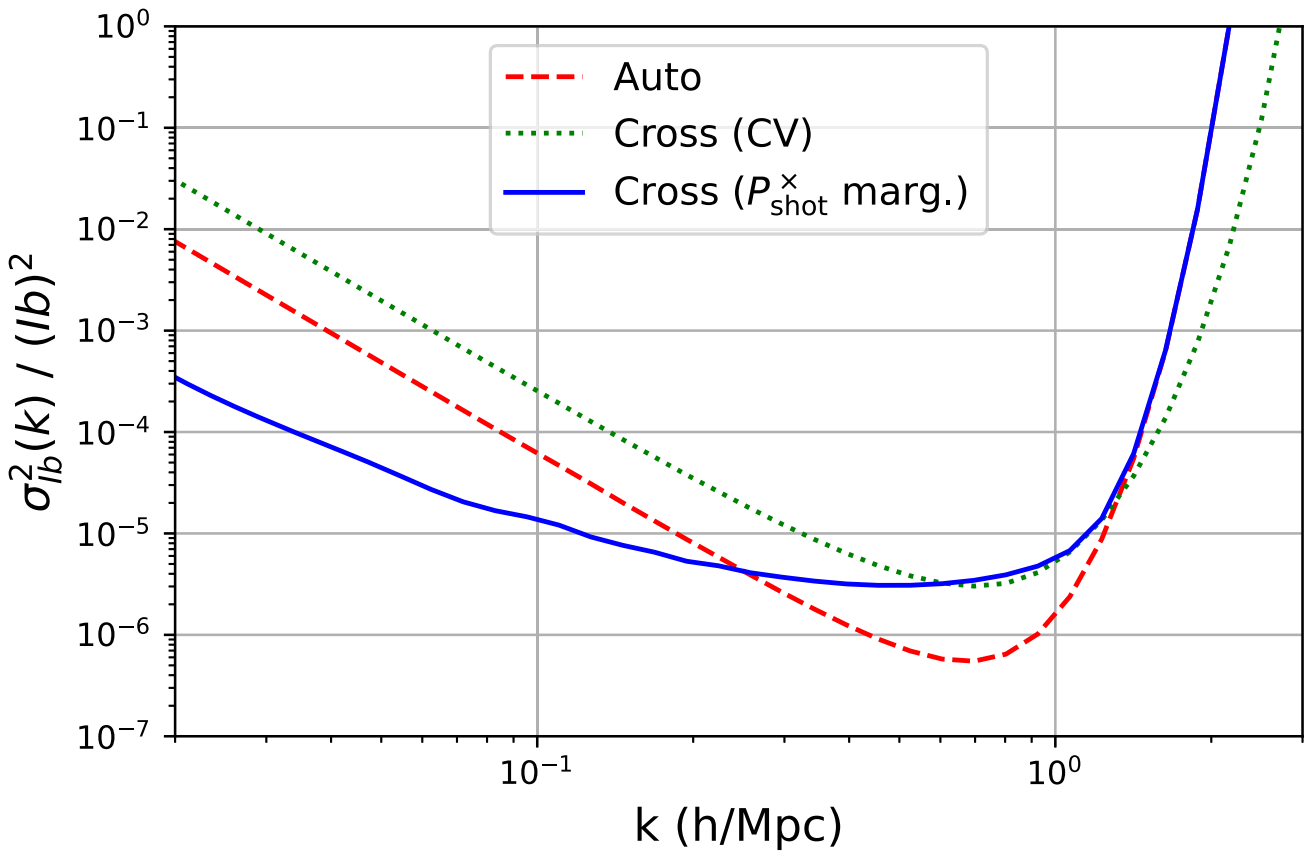

FIG. 4. Fractional variance per mode in $[Ib]$. We model errors in the intensity map autocorrelation, and cross-correlation marginalized over $P^\times_{\rm shot}$, specified by Equations (17) and (20), respectively. The instrument design is specified in Section III B.

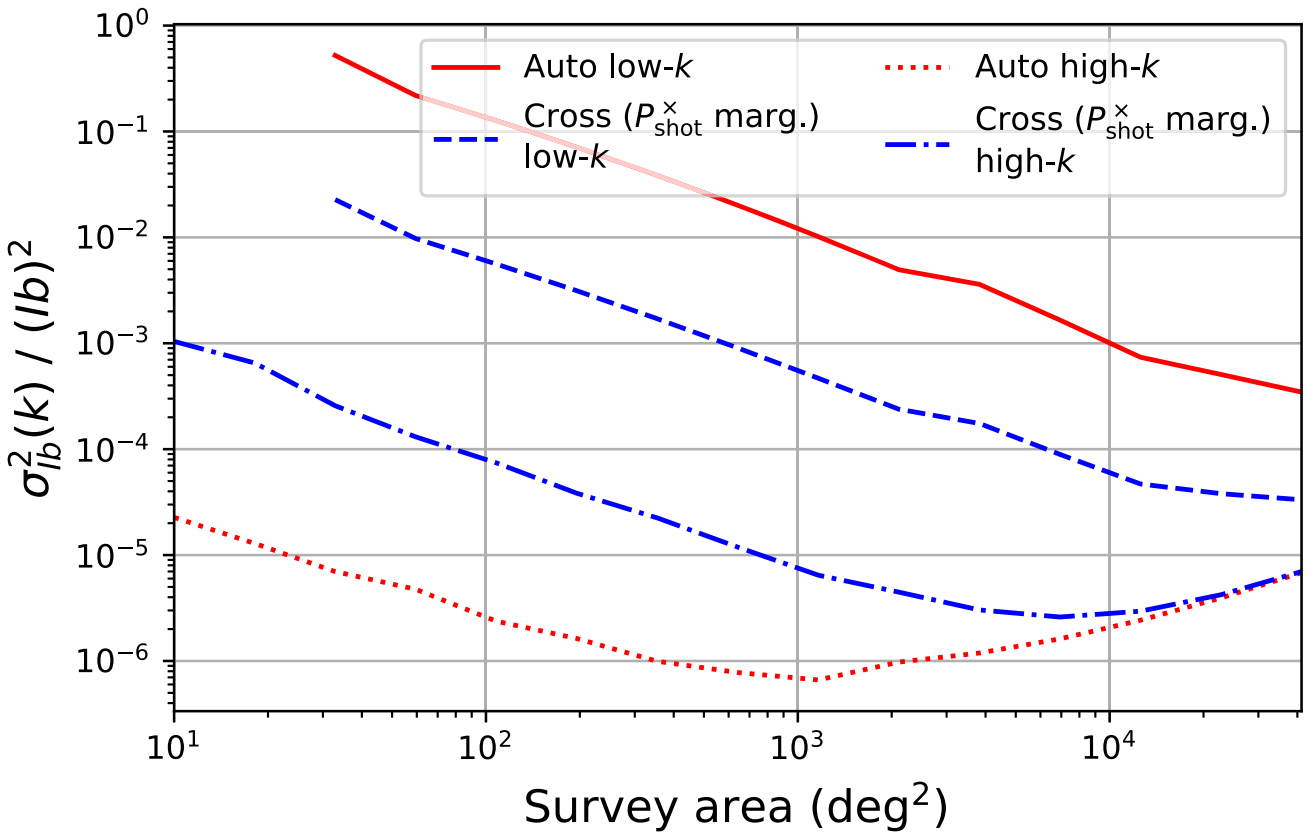

FIG. 5. Fractional variance per mode in $[Ib]$ varying the survey area from 10 to 40000 deg$^2$ (full-sky) at fixed survey time.

For $k = 0.80\ \mathrm{h\,Mpc^{-1}}$ modes, the autopower produces smaller errors than the cross power for almost all survey depths. Here, noise in the galaxy redshift survey exceeds cosmic variance, i.e., $W_g > 1$. For the largest survey areas, however, the crosspower performs similarly to the autopower. Though we marginalize over the cross-shot power, care must be taken to understand nonlinear physics on these scales to validate this conclusion. The autocorrelation variance for $k = 0.80\ \mathrm{h\,Mpc^{-1}}$ is minimized for $A_{\rm surv} \approx 1100\ \mathrm{deg}^2$. This corresponds to the survey depth that matches cosmic variance, i.e., $W_I \approx 1$, matching conclusions described in Knox [14]. The cross-correlation variance features a similar minimum instead at $A_{\rm surv} \approx 10,000\ \mathrm{deg}^2$. On this scale, the depth of the intensity map matches that of the galaxy redshift survey, i.e., $W_I \approx W_g$. Finally, we note that in the case of increased noise in the intensity map, these minima would be pushed to proportionally smaller survey areas.

## IV. CONCLUSION

Intensity mapping cross-correlations provide a powerful means to measure the biased intensity of a given tracer. Cross-correlation has well-established utility to reject sources of contamination in the intensity map that are not correlated with the target galaxy population, such as interlopers and foregrounds [8–10]. Furthermore, cross-correlation measurements of the biased intensity can evade cosmic variance.

For a given $k$ mode, cosmic variance can be mitigated if the fractional noise in the clustering power spectrum in both surveys (i.e., $W_I$ and $W_g$) is less than 1, and if the stochasticity between the surveys is small. These requirements are straightforwardly achieved on large, linear scales. Smaller scales tend to feature higher fractional noise and higher stochasticity due to the dominance of nonlinear physics. On smaller scales, an intensity mapping autocorrelation may be preferable to a cross-correlation, because the latter introduces an additional source of noise from the cross-correlating survey, which may exceed cosmic variance. Note, however, that the use of the autocorrelation requires stronger control of contamination from foregrounds and interlopers. In either case, the depth of the intensity map is optimized when fractional noise in the intensity map matches that of the cross-correlating survey. We stress that this optimal depth depends on the scales of interest and the level of noise in the two surveys. There is therefore no one-size-fits-all optimum survey depth, but careful analyses using the techniques and equations described in this paper provide a useful framework for designing a survey to achieve specific science goals.

## ACKNOWLEDGMENTS

We thank Patrick Breysse and Anthony Pullen for useful discussion about LIM cross-correlations and careful review of the draft, and David Leisawitz for discussion on the midinfrared to far-IR photon background. T. M. O. especially thanks Peter Timbie for invaluable guidance and advising. We also thank our reviewer for useful comments. T. M. O. thanks the Wisconsin Space Grant for fellowship support, the National Space Grant and Universities Space Research Association for internship support, and the UW-Madison graduate school for training and support. We acknowledge the use of the Astropy, COLOSSUS [32], Matplotlib, and NumPy PYTHON modules.